\documentclass[%
reprint,
superscriptaddress,
showpacs,preprintnumbers,
amsmath,amssymb,
aps,
prb,
dvipdfm
]{revtex4-1}
\usepackage{graphicx}
\usepackage{dcolumn}
\usepackage{bm}
\usepackage{amsmath}	
\begin{document}

\newcommand{\BigFig}[1]{\parbox{12pt}{\Huge #1}}
\newcommand{\BigZero}{\BigFig{0}}

\renewcommand{\topfraction}{0.99}
\renewcommand{\bottomfraction}{0.99}
\renewcommand{\dbltopfraction}{0.99}
\renewcommand{\textfraction}{0.1}
\renewcommand{\floatpagefraction}{0.99}
\renewcommand{\dblfloatpagefraction}{0.99}

\def\U#1{{%
\def\O{\mbox{O}}
\def\u{\mbox{u}}
\mathcode`\u=\mu
\mathcode`\O=\Omega
\mathrm{#1}}}
\def\ii{{\mathrm{i}}}
\def\jj{\,\mathrm{j}}                   
\def\ee{{\mathrm{e}}}
\def\dd{{\mathrm{d}}}
\def\Re{\mathop{\mathrm{Re}}}
\def\Im{\mathop{\mathrm{Im}}}
\def\bra#1{\langle #1|}
\def\braa#1{\langle\langle #1|}
\def\ket#1{|\mbox{$#1$}\rangle}
\def\kett#1{|\mbox{$#1$}\rangle\rangle}
\def\bracket#1{\langle\mbox{$#1$}\rangle}
\def\bracketi#1#2{\langle\mbox{$#1$}|\mbox{$#2$}\rangle}
\def\bracketii#1#2#3{\langle\mbox{$#1$}|\mbox{$#2$}|\mbox{$#3$}\rangle}
\def\vct#1{{\mathchoice{\mbox{\boldmath$#1$}}{\mbox{\boldmath$#1$}}%
  {\mbox{\scriptsize\boldmath$#1$}}{\mbox{\scriptsize\boldmath$#1$}}}}
\def\fracpd#1#2{\frac{\partial#1}{\partial#2}}
\def\rank{\mathop{\mathrm{rank}}} 
\def\sub#1{_{\mbox{\scriptsize#1}}}
\def\sur#1{^{\mbox{\scriptsize#1}}}
\def\kagome{kagom\'{e} }
\def\Kagome{Kagom\'{e} }
\def\kagometype{kagom\'{e}-type }
\def\Kagometype{Kagom\'{e}-type }

\title{Observation of flat band for terahertz spoof plasmon\\ in metallic \kagome lattice}

\author{Yosuke Nakata}
\email{nakata@giga.kuee.kyoto-u.ac.jp}
\affiliation{Graduate School of Engineering, Kyoto University, Kyoto 615-8510, Japan}
\author{Takanori Okada}
\affiliation{Pioneering Research Unit for Next Generation, Kyoto University, Kyoto 611-0011, Japan}
\author{Toshihiro Nakanishi}
\affiliation{Graduate School of Engineering, Kyoto University, Kyoto 615-8510, Japan}
\author{Masao Kitano}
\email{kitano@kuee.kyoto-u.ac.jp}
\affiliation{Graduate School of Engineering, Kyoto University, Kyoto 615-8510, Japan}

\date{\today}

\begin{abstract}
We study the dispersion relation of a
metamaterial composed of metallic disks and bars arranged
to have \kagome symmetry and find that a plasmonic flat 
band is formed by the topological nature of the \kagome lattice.
To confirm the flat-band formation,
we fabricate the metamaterial and make transmission measurements
in the terahertz regime. 
Two bands formed by transmission minima
that depend on the polarization of the incident terahertz beams are observed.
One of the bands corresponds to the flat band, as confirmed by the
fact that the resonant frequency is almost independent of
the incident angle.
\end{abstract}

\maketitle

 \section{Introduction}

\Kagome lattices have attracted
considerable interest from the aspect of geometric frustration in condensed-matter physics.\cite{atwood_kagom|[eacute]|_2002}
There is extensive degeneracy of nondispersing resonant modes in a resonator system with \kagome symmetry.
These eigenmodes form a flat band, where the resonant frequency of the band is the same for all wavevectors in the first Brillouin zone. The flat band originates purely from the topology of the lattice structure, and it remains flat even when the couplings between the adjacent sites are significantly large. 
Furthermore, such flat bands can lead to 
ferromagnetism of itinerant fermions,\cite{lieb_two_1989,mielke_ferromagnetic_1991,mielke_ferromagnetism_1991,tasaki_ferromagnetism_1992} 
supersolidity for bosons,\cite{huber_bose_2010,moeller_correlated_2012}
crystalline ordering,\cite{wu_flat_2007} and other effects.

Although the flat-band formation is first expected in quantum systems, it can occur in electromagnetic systems.
The presence of electromagnetic flat bands in \kagome lattices has
already been predicted theoretically in some electromagnetic systems, 
such as two-dimensional photonic crystals\cite{takeda_flat_2004} and 
metallophotonic waveguide networks.\cite{endo_tight-binding_2010}
In the flat band, the group velocity is 
slowed down in all directions
and the effective mass of the photons becomes very heavy.
It is important to study the flat band in 
the electromagnetic system with \kagome symmetry
in terms not only of fundamental physics, 
but also from an application standpoint, such as slow light;\cite{baba_slow_2008} 
however, there has been no experimental demonstration for the electromagnetic flat band.

Here, we focus on the flat band for a terahertz (THz) plasmonic mode.
Although there is no surface plasmon of metals in the THz region, 
the modes having the dispersion relation 
similar to surface plasmons are formed 
in structured metals,
and called spoof surface plasmons.\cite{pendry_mimicking_2004,garcia-vidal_surfaces_2005,williams_highly_2008,maier_terahertz_2006}
In this paper, 
we theoretically and experimentally obtain the 
dispersion relation for a spoof plasmon in the metallic \kagome lattice and 
demonstrate the electromagnetic flat band in the THz regime. 
Numerical simulations are also performed to
provide confirmation of the experiments.

\section{Theoretical model}
\begin{figure}[!b]
\includegraphics[width=8.5cm]{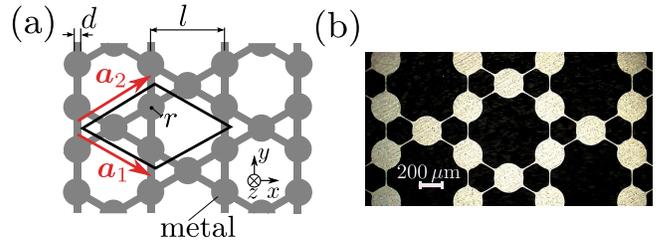}
\caption{\label{fig:kbdrs}(Color online) (a) Schematic view of \kagometype bar-disk resonators (KBDRs).
(b) The fabricated KBDRs on a stainless-steel plate with $l=800$,
$d=10$, and $r=145$, and thickness $h=30\,\U{\mu
 m}$.
}
\end{figure}
We introduce \kagometype bar-disk 
resonators (KBDRs), shown in 
Fig.~\ref{fig:kbdrs}(a).
Metallic disks and narrow bars are connected to form a \kagome lattice.
KBDRs are artificially engineered metallic structures, and considered as a kind of metamaterial.\cite{veselago_electrodynamics_1968,shelby_experimental_2001,pendry_magnetism_1999,pendry_negative_2000,leonhardt_optical_2006,pendry_controlling_2006,schurig_metamaterial_2006,liu_coupled_2009}
In KBDRs, electric charge stored on each disk 
temporally oscillates between positive and negative values.
We discuss the formation of a flat band in KBDRs by 
using a coupled oscillator model.
We denote the electric potential at the $i$th disk as $\phi_i$. 
Introducing $\Phi_i:=\int \phi_i \dd t$,
we obtain a Lagrangian $\mathcal{L}$ as
\begin{equation}
 \mathcal{L}= \frac{C}{2} \sum_{i} \dot{\Phi}_i^2 + C\sub{M} \sum_{i,j} \frac{A_{ij}}{2}\dot{\Phi}_i\dot{\Phi}_j-
\frac{1}{2L}\sum_{i,j} \frac{A_{ij}}{2}(\Phi_i-\Phi_j)^2,  \label{eq:1}
\end{equation}
with capacitance $C$ of the disk, inductance $L$ of the bar, 
coefficient of electric induction $C_\mathrm{M}$ between nearest disks,
and adjacency matrix $A_{ij}$ of the \kagome lattice, whose element is $1$ if the  $i$th and $j$th disks
are directly connected by a bar for $i\ne j$; otherwise 0.\cite{biggs_algebraic_1994}
The first, second, and third terms of Eq.~(\ref{eq:1}) 
represent the electric energy stored on disks, 
mutual electric energy stored between disks,
and magnetic energy stored around bars, respectively.
Here, we consider only the nearest mutual couplings.

\begin{figure}[!t]
\includegraphics[width=8.5cm]{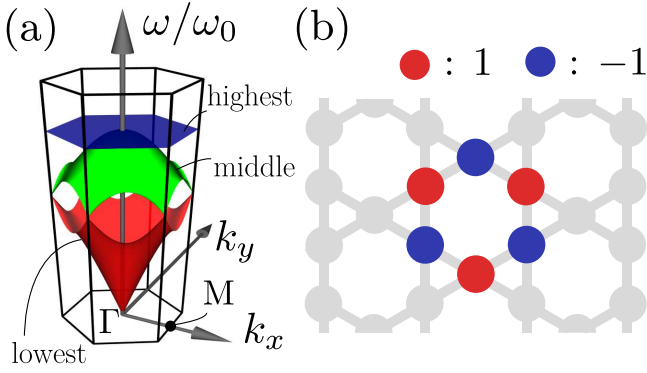}
\caption{\label{fig:theoretical_band}(Color online) 
(a) Dispersion relation of KBDRs for $\kappa=0$.
(b) Localized eigenmode of KBDRs. 
}
\end{figure}

Using the Euler-Lagrange equation 
$({\dd}/{\dd t})({\partial \mathcal L}/{\partial \dot{\Phi}_i})- {\partial \mathcal L}/{\partial \Phi_i}=0$,
we obtain coupled charge equations as
\begin{equation}
 \ddot{q}_i + \omega_0^2\big(4q_i-\sum_jA_{ij} q_j\big)+\kappa \sum_jA_{ij}\ddot{q}_j=0, \label{eq:2}
\end{equation}
with stored charge $q_i=C\dot{\Phi}_i$ at the $i$th disk, resonant angular frequency $\omega_0=1/\sqrt{LC}$, and coupling coefficient $\kappa=C\sub{M}/C$.
In the frequency domain, we rewrite Eq.~(\ref{eq:2}) as
\begin{equation}
\sum_j A_{ij} \tilde{q}_j = \frac{4-(\omega/\omega_0)^2}{1+\kappa(\omega/\omega_0)^2} \tilde{q}_i,  \label{eq:3}
\end{equation}
where tildes represent complex amplitudes.
Owing to the lattice symmetry, we can reduce Eq.~(\ref{eq:3})
to an eigenvalue problem for a $3\times3$ matrix 
and obtain the dispersion relation consisting of three bands as
\begin{equation}
 \frac{\omega}{\omega_0} = \sqrt{\frac{6}{1-2\kappa}},\ 
\sqrt{\frac{3+2(3+F)\kappa\pm(1+4\kappa)\sqrt{3+2F}}{1+2\kappa-2(1+F)\kappa^2}},  \label{eq:4}
\end{equation}
where $F=\cos \vct{k}_\parallel\cdot \vct{a}_1+\cos \vct{k}_\parallel\cdot \vct{a}_2+\cos \vct{k}_\parallel\cdot (\vct{a}_1-\vct{a}_2)$ with wavevector $\vct{k}_\parallel$ in the $xy$ plane and unit-lattice vectors 
$\{\vct{a}_1,\ \vct{a}_2\}$ shown in Fig.~\ref{fig:kbdrs}(a).
A calculated dispersion relation is shown in Fig.~\ref{fig:theoretical_band}(a) for $\kappa=0$. 
It is clear that the highest band $\omega/\omega_0=\sqrt{6/(1-2\kappa)}$ is flat or independent of $\vct{k}_\parallel$. 
It can be seen that the lowest band shows 
conical dispersion near the $\Gamma$ point
and that the bending middle band touches the flat band at the $\Gamma$ point.

The flat band is 
caused by the interference of spoof plasmon propagating in the \kagome lattice.
The adjacency matrix $A_{ij}$ of the \kagome lattice 
has eigenmodes localized at hexagonal sites with an eigenvalue of $-2$. 
One of them is shown in Fig.~\ref{fig:theoretical_band}(b). 
The number of the eigenmodes is equal to the number of hexagons in the \kagome lattice.
The flat band is formed from these degenerated localized modes 
as they are not coupled with each other.
\begin{figure}[!b]
\includegraphics[width=8.5cm]{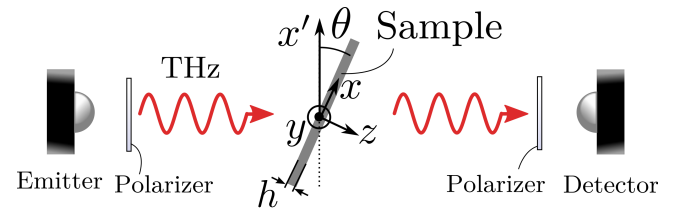}
\caption{\label{fig:experiment}(Color online) 
Schematic view of the transmission experiment. 
The plane of a sample is represented by the coordinate 
system $(x,y)$ shown in Fig.~\ref{fig:kbdrs}(a).
}
\end{figure}

\section{Experimental setup}
We fabricate KBDRs on a stainless-steel plate.
The dimensions depicted in Fig.~\ref{fig:kbdrs}(a)
are as follows:
period between bars $l=800\,\U{\mu m}$,
bar width $d=10\,\U{\mu m}$,
disk radius $r=145\,\U{\mu m}$,
and metal thickness $h=30\,\U{\mu m}$.
The size of the  area patterned KBDRs is $1.1\,\U{cm}\times1.1\,\U{cm}$.
A photomicrograph of a fabricated sample 
is shown in Fig.~\ref{fig:kbdrs}(b).

To investigate the dispersion relation experimentally,
we perform THz time-domain spectroscopy (THz-TDS), shown in Fig.~\ref{fig:experiment}.
A THz emitter and detector (EKSPLA Ltd.) with dipole antennas
attached to Si lenses are used. These antennas are integrated 
on low-temperature-grown GaAs photoconductors,
and driven by a femtosecond fiber laser (F-100, IMRA America, Inc.) with
a wavelength of $810\,\U{nm}$ and pulse duration of $120\,\U{fs}$.
The THz beam is collimated with the Si lens near the emitter. The beam
radius is about $3.7\,\U{mm}$, which covers a large number of KBDRs.
The THz electric field $E(t)$ is coherently measured
by the detector. We obtain the transmission 
spectrum $T(\omega)$ in the frequency domain from 
$T(\omega)=|\tilde{E}\sub{s}(\omega)/\tilde{E}\sub{ref}(\omega)|^2$,
where $\tilde{E}\sub{s}(\omega)$ and $\tilde{E}\sub{ref}(\omega)$
are Fourier transformed electric fields with and without the sample, 
respectively.

In order to obtain the band structure between the $\Gamma$ point and the $\mathrm{M}$ point,
the sample is rotated by $\theta$ with respect to the $y$ axis from normal incidence
(Fig.~\ref{fig:experiment}).
The angles $\theta$ range from $\theta=0^\circ$ to $\theta=65^\circ$
with a step $\Delta\theta = 2.5^\circ$.
The magnitude of the wavevector $\vct{k}_\parallel$ on the sample plane is
given by $k_\parallel=(\omega/c) \sin\theta,$
where $c$ is the speed of light.

We perform transmission experiments 
for two polarizations along the $x'$ axis (parallel configuration) 
and $y$ axis (perpendicular configuration). 
We denote the electric field of the incident wave as $\vct{E}$,
and the projection of $\vct{E}$ to the sample plane as $\vct{E}_\parallel$. 
For parallel or perpendicular configurations, 
$\vct{E}_\parallel$ is parallel or perpendicular, 
respectively, to $\vct{k}_\parallel$.
Wire-grid polarizers near the emitter and detector are adjusted
so that the emitted and detected fields have the same polarization.

\section{Results}
\begin{figure}[!tb]
 \includegraphics[width=8.3cm]{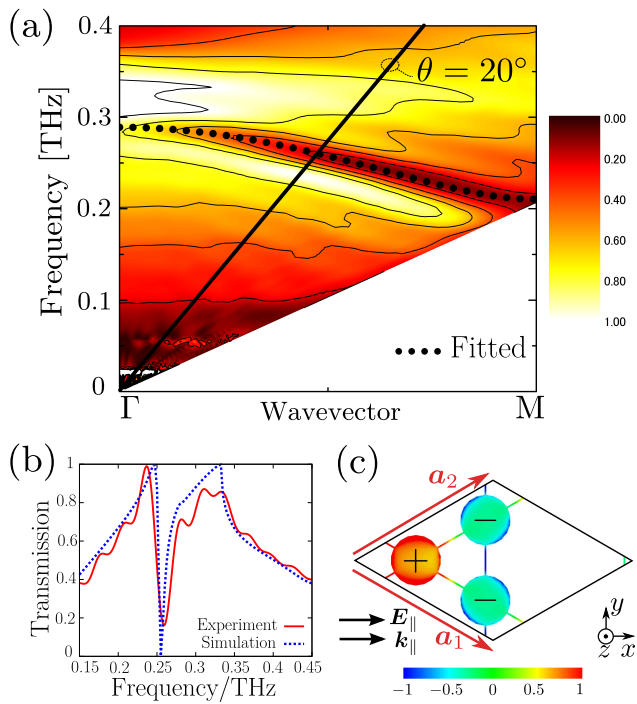}
\caption{\label{fig:results_para4}
(Color online) Parallel configuration ($\vct{E}_\parallel \parallel \vct{k}_\parallel$). 
 (a) Experimentally obtained transmission diagram of KBDRs. 
Transmission minima between $0.21\,\U{THz}$ and $0.28\,\U{THz}$
are observed and theoretically fitted by a dotted line.
The solid line corresponds to $\theta=20^\circ$.
(b) Transmission spectrum for $\theta=20^\circ$ obtained by simulation and experiment.
(c) Surface electric charge distribution obtained by simulation
at the transmission minimum $0.255\,\U{THz}$ for $\theta=20^\circ$ in the unit cell.
}
\end{figure}

Figure~\ref{fig:results_para4}(a) 
displays the transmission spectrum for parallel configuration. 
The wavevectors are estimated as $k_\parallel=(\omega/c) \sin\theta$.
Transmission spectrum minima are observed 
from $0.21\,\U{THz}$ to $0.28\,\U{THz}$.
With an increase of wavenumber,
the frequency of the transmission minimum decreases 
from $0.28\,\U{THz}$ at the $\Gamma$ point
and approaches $0.21\,\U{THz}$ at the $\mathrm{M}$ point.

For further investigation,
we calculate the electromagnetic 
response of KBDRs for parallel configuration.
A commercial 
finite-element method solver (Ansoft HFSS) is used.
In the simulation, a plane THz wave is injected into perfectly conducting
KBDRs at an incident angle $\theta$. 
By using periodic boundaries with some phase shifts, 
the transmission and the electromagnetic fields in the unit cell
are calculated for an oblique incident plane wave.
The measured transmission spectra for $\theta =20^\circ$ are compared with the simulation in Fig.~\ref{fig:results_para4}(b).
The frequency of the transmission minimum and shape of the curve for the simulation are consistent with the experimental result, 
which confirms the validity of the assumption of perfect conductors.

Figure~\ref{fig:results_para4}(c) shows the calculated distribution of 
surface electric charges 
at a minimum ($0.255\,\U{THz}$) for $\theta=20^\circ$.
This mode corresponds to the middle band.
Disks along the $x$ axis are alternately charged.
The in-phase currents along $\vct{a}_1$ and $\vct{a}_2$ are
excited by the electromotive 
force due to $\vct{E}$. 
No resonance appears for $\theta=0$ 
because the current flowing into a disk is balanced
by the current flowing out of it.
For the excitation of this mode, 
a phase-shifted field in the $x$-direction is needed.

By using Eq.~(\ref{eq:4}), the fitting parameters are obtained 
from experimental data as 
$\omega_0/(2\pi)=0.105\,\U{THz}$ and $\kappa=0.103$.
The resultant dispersion curve is represented as a dotted curve in 
Fig.~\ref{fig:results_para4}(a).
Positive charges on one disk induce
negative charges on the other; therefore, 
$\kappa<0$ is ordinarily expected in the static limit ($\omega\rightarrow 0$). 
It seems strange that $\kappa$ would be positive.
In our situation, it can be explained by
a retardation effect.\cite{solymar_waves_2009}
The phase shift between nearest disks is given by
$(\omega_c/c)\times l/\sqrt{3} \sim 0.77\times \pi$ 
at the center frequency $\omega_c/(2\pi)=0.25\,\U{THz}$ 
of the middle band. The near $\pi$ shift leads to $\kappa>0$.
Although $\kappa$ depends on frequency,
we can approximately regard it as a constant 
between $0.2\,\U{THz}$ and $0.3\,\U{THz}$.

\begin{figure}[!tbhp]
 \includegraphics[width=8.3cm]{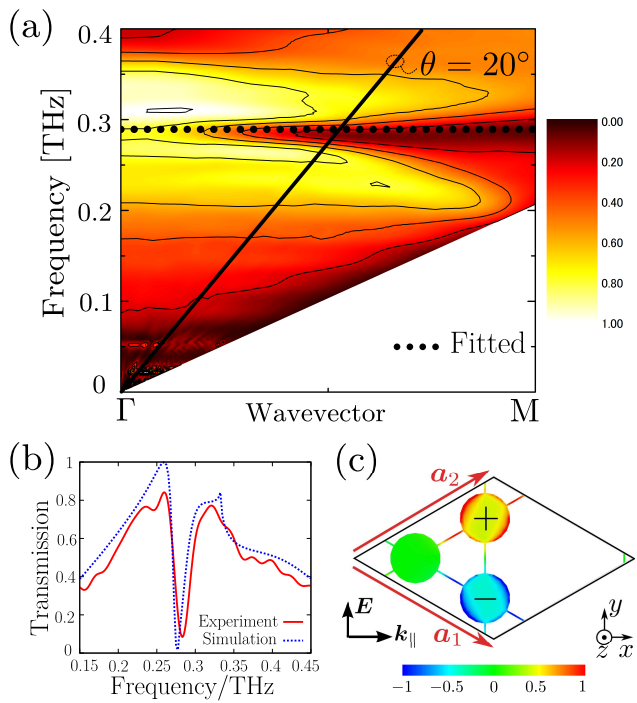}
\caption{\label{fig:results_perp4}
(Color online) Perpendicular configuration ($\vct{E} \perp \vct{k}_\parallel$).
 (a) Experimentally obtained transmission diagram of KBDRs. 
A flat transmission band 
is observed around $0.28\,\U{THz}$ and 
theoretically fitted by the dotted line.
The solid line corresponds to $\theta=20^\circ$.
(b) Transmission spectrum for $\theta=20^\circ$ obtained by simulation and experiment.
(c) Surface electric charge distribution obtained by simulation
at a transmission minimum of $0.278\,\U{THz}$ for $\theta=20^\circ$ 
in the unit cell.
}
\end{figure}

Figure~\ref{fig:results_perp4}(a) 
displays the transmission spectrum for perpendicular configuration.
Unlike in the case of parallel configuration, the flat band of the transmission minima is observed around $0.28\,\U{THz}$.

In order to confirm that the flat band is due to the interference of a spoof plasmon, we perform a simulation for perpendicular configuration.
A calculated transmission spectrum by simulation for $\theta=20^\circ$ is shown in
Fig.~\ref{fig:results_perp4}(b) with the experimental data.
We see a good agreement in the frequency of the transmission minimum and 
the shape of the curve.
The calculated distribution of surface 
electric charges at a minimum ($0.278\,\U{THz}$) 
for $\theta=20^\circ$ is shown in Fig.~\ref{fig:results_perp4}(c).
The resonant mode has antisymmetric amplitudes on the right two disks and
there is no charge stored on the left disk.
This mode can be constructed by the localized modes shown in Fig.~\ref{fig:theoretical_band}(b).
Therefore, the flat transmission minima are caused by the topological 
nature of the \kagome lattice.
The mode is excited by anti-phase electromotive 
force caused by $\vct{E}$ along bars parallel to $\vct{a}_1$ and $\vct{a}_2$.
The electromotive force along vertical bars 
does not contribute to the storage of charges on disks because
the currents flowing into and out of a disk are balanced.
In the case of $\theta=0$, the current flowing into a disk is equal 
to the current flowing out of it and the flat-band mode cannot be excited.

A dotted line in Fig.~\ref{fig:results_perp4}(a) represents the highest band given by Eq.~(\ref{eq:4}) with the previously derived parameters $\omega_0/(2\pi)=0.105\,\U{THz}$ and $\kappa=0.103$.
It fits well with the minima experimentally obtained.
The bend of the flat band caused by coupling to second (or higher) nearest sites
is negligibly small, so the assumption of only nearest mutual disk coupling is appropriate.

\section{discussion}

Our bar-disk resonators (BDRs) are 
obtained by inverting the metallic area and empty space of the 
slit-hole resonators (SHRs)\cite{liu_extraordinary_2009,zhu_electric_2010} composed of slits and holes engraved on an ultra thin metallic plate.
The BDRs and SHRs are complementary structures related through the Babinet's principle,\cite{jackson_classical_1999, born_principles_1999,falcone_babinet_2004}
based on the electric/magnetic reciprocity of a vacuum.
We denote a pair of an electric field $\vct{E}$ and a 
magnetic field $\vct{H}$ as $(\vct{E},\vct{H})$.
Due to the Babinet principle, the transmittance of 
a complementary metallic screen with
an complementary incident wave $(\vct{E}',\vct{H}')=(Z_0\vct{H},-\vct{E}/Z_0)$
is equal to the reflectance of the original metallic screen
illuminated by an incident wave $(\vct{E},\vct{H})$,
where $Z_0$ is the impedance of a vacuum.
Thus, the transmission peaks in SHRs 
correspond to the transmission minima in BDRs.
This fact shows the duality of the Lagrangians of SHRs and BDRs.

Electromagnetic flat bands for all crystal directions 
have been reported for photonic crystals
with square symmetry, theoretically\cite{altug_two-dimensional_2004} and experimentally.\cite{altug_experimental_2005}
In this case, the flat bands are formed due to
good lateral confinement (high $Q$ factor) of the quadropole modes,
which lack preferential coupling directions,
at defects of photonic crystals.
On the other hand, the flat band for KBDRs is not caused 
by highly confined modes,
but by the topological nature of the \kagome lattice.
The \kagome lattice prevents
spoof plasmons from propagating despite the existence of strong coupling.
Thus, the physical origin of the flat band on KBDRs 
differs from that for photonic crystals.\cite{altug_two-dimensional_2004, altug_experimental_2005}
The flat band for propagating modes has been 
theoretically predicted 
for square waveguide networks.\cite{feigenbaum_resonant_2010}
Our system is considered as an experimental 
realization of the flat band for the propagating mode.

The flat band in the \kagome lattice comes from local interference effects.
The global symmetry (i.e., periodicity of the lattice) is 
not necessarily required because local symmetries can support 
the localized mode.  The resonance independent of the 
incident angle could be expected for the metallic structure 
having localized modes with the same resonant frequency,
even without periodicity.

\section{conclusion}

In conclusion, we studied theoretically and experimentally 
the electromagnetic flat band on a metallic \kagome lattice. 
\Kagometype bar-disk resonators were proposed to realize the flat band.
A dispersion relation composed of three bands was theoretically predicted
for KBDRs. The highest band was flat for all wavevectors.
Two bands formed by transmission minima depending on the polarization of 
the incident terahertz beams were observed experimentally. 
One of the bands corresponded to the flat band.
Theoretical fitting showed good agreement for these modes.
By simulation, we revealed that the flat band was caused by
the topological nature of the \kagome lattice.

The flat band can be applicable to slow light,  
which is useful for the control of group velocity,\cite{Tamayama_2010,Tamayama_2012} 
high sensitive sensing,
and other applications.
In the flat band, the effective mass of photons becomes very heavy
and their correlation has an important role.
Multiphoton correlation effects in 
\kagome lattices are important in terms of fundamental physics and 
should be studied in the future. 

\begin{acknowledgments}
The authors would like to thank
A. Yao and T. Hikihara for technical assistance,
and S. Endo for fruitful discussions.
This research was supported in part by Grants-in-Aid for Scientific Research No.~22109004 and No.22560041,
the Global COE program ``Photonics and Electronics Science and Engineering'' at Kyoto University,
the Program for Improvement of Research Environment for Young Researchers from the Special Coordination Funds (SCF) for Promoting Science and Technology  commissioned by the Ministry of Education, Culture, Sports, Science and Technology (MEXT) of Japan (T.O.), the Research Foundation for Opto-Science and Technology (T.O.),
and research grants from 
the Murata Science Foundation (T.O. and T.N.).
\end{acknowledgments}

\nocite{*}

%

\end{document}